\documentclass[10pt]{iopart}
\usepackage[utf8]{inputenc}
\usepackage[square,sort&compress,numbers]{natbib}
\expandafter\let\csname equation*\endcsname\relax
\expandafter\let\csname endequation*\endcsname\relax
\usepackage{amsmath,amsfonts,graphicx,mathrsfs,amssymb,multirow}
\usepackage[colorlinks=true]{hyperref}
\usepackage{outlines}

\renewcommand{\thefootnote}{\arabic{footnote}}

\begin{document}

\title{An Exact, Coordinate Independent Classical Firewall Transformation}

\author{Nathaniel A. Strauss, Bernard F. Whiting }
\address{Department of Physics, University of Florida, Gainesville, Florida, USA}
\eads{\mailto{straussn@ufl.edu}, \mailto{bernard@phys.ufl.edu}}

\date{\today}

\begin{abstract}
A proposal for resolving the black hole information paradox was recently put forward by 't Hooft in the form of his firewall transformation.  Although this proposal has begun to gain some limited traction, its physical foundation is still somewhat obscure.  Here we develop a classical Hamiltonian analog, which is oriented towards quantization, by using the canonical formalism developed by Arnowitt, Deser, and Misner (ADM).  We use a model of two null, spherical shells in a Schwarzschild black hole background, and within our ADM formalism we are able to characterize the dynamics of the entire system, especially at the point of collision, and we reproduce the related Dray-'t Hooft-Redmount formula.
Finally, we are able to find a classical analog for 't Hooft's firewall transformation. Unlike 't Hooft's firewall transformation and previous classical analogs, the classical firewall transformation we obtain is free from approximation and maintains the coordinate independence of the ADM formalism. We leave to future work the quantization of the theory. 
\end{abstract}
\noindent{\it Keywords\/}: general relativity, black hole, quantum mechanics, firewall, information

\maketitle

\renewcommand{\thefootnote}{\roman{footnote}}
\section{Introduction}

Due to the lack of a complete quantum theory of gravity, physical questions about the collapse, the singularity, and the event horizon of black holes have been probed by the use of simplified classical models that admit a straightforward quantization procedure, usually using the group quantization method, a generalization of Dirac's quantization procedure.~\citep{Hajicek_1995, Hajicek_1995_2} A common simplified model is a spherically symmetric distribution of matter collapsing or expanding in a Schwarzschild black hole background, e.g.~\citep{stephens_1994,Parentani:1994ij,Ambrus_2005,Vaz_2001,Leal_2017,Vaz_2022_2,Vaz_2022,Hajicek_1999,Hajicek_2000,Bicak_2003,Fiamberti_2007,Corichi_2001,Campiglia_2016,Menotti_2009,Menotti_2010,Eyheralde_2017,Eyheralde_2019}. For example, quantization procedures for a single shell collapse demonstrate completely unitary evolution as the shell collapses and subsequently bounces back outwards after a long black-hole-like epoch.~\citep{Hajicek_1999,Hajicek_2000} Similar results have been shown for models including two or more shells.~\citep{Bicak_2003,Fiamberti_2007} Further analyses suggest that the returning wavefunction could be expanding in a new spacetime through the singularity.~\citep{Corichi_2001,Campiglia_2016}   Shell models have also been used as geometrical backgrounds to Hawking radiation. In~\citep{Menotti_2009,Menotti_2010} the shells were used to refine the description of Hawking radiation to include the changing mass of the black hole, and in~\citep{Eyheralde_2017,Eyheralde_2019} there is theoretical evidence that at least some quantum information about an ingoing shell crossing the event horizon is recoverable from Hawking radiation, which may help solve the black hole information paradox.

The black hole information paradox remains a central problem in the simultaneous application of quantum field theory and general relativity.~\citep{mathur_2009} In recent years, Gerard 't Hooft has put forward a theoretical framework intended to resolve the information paradox via a so-called ``firewall transformation'', which preserves, in the form of changes in coordinates of outgoing particles, the information usually lost by particles falling through a black hole's event horizon.~\citep{'thooft_2016, 'thooft_2016_2,'thooft_2016_3,'thooft_2018,'thooft_2021,'thooft_2022,'tHooft_2023} The theory has gained some traction in the search for black hole microstates~\citep{hogan_2020,zeng_2021,kwon_2022,slagter_2022,egorov_2022} and is conceptually similar to Hawking radiation models developed on quantum shell backgrounds, e.g.~\citep{Eyheralde_2017,Eyheralde_2019}. The essence of the firewall transformation can be formulated as an invertible change in basis between the quantum observables for the ingoing and outgoing particles:~\citep{'thooft_2018}
\begin{align}
    \hat{u}_\text{in}(\Omega_j)&=\sum_i f(\Omega_j,\Omega_i) \hat{p}_\text{out}(\Omega_i),\label{eq:firewall_thooft_1}\\
    \hat{u}_\text{out}(\Omega_j)&=\sum_if(\Omega_j,\Omega_i) \hat{p}_\text{in}(\Omega_i),\label{eq:firewall_thooft_2}
\end{align}
where $\hat{u}_\text{in/out}$ and $\hat{p}_\text{in/out}$ represent an ingoing or outgoing particle's quantum position and momentum operators (in Kruskal coordinates), respectively, $\Omega_j$ and $\Omega_i$ represent the angular position of the particles relative to the black hole, and $f(\Omega_j,\Omega_i)$ gives the shift in particle $j$'s Kruskal position operator due to particle $i$. Equations~\eqref{eq:firewall_thooft_1} and~\eqref{eq:firewall_thooft_2} are quantizations of the classical result that a null particle's constant Kruskal coordinate will change when it passes another null particle, which has recently been recognized as a kind of Shapiro time-delay effect.~\citep{'thooft_2022,'tHooft_2023}

The quantum firewall transformation proposed by 't Hooft uses a Hamiltonian framework for the particle dynamics in the black hole background, which is necessary in a quantum problem. However, the gravitational interaction between the particles resulting in the Shapiro time delay, which forms the basis for the firewall transformation, is derived in a non-Hamiltonian framework. The insertion of a non-Hamiltonian result into a Hamiltonian theory before quantization obscures the meaning of the resulting quantum variables. This paper aims to contribute to bridging the gap between the two aforementioned bodies of literature: the semiclassical and quantum models for spherical shells in a black hole background and the firewall transformation as proposed by 't Hooft. We derive the classical expressions analogous to~\eqref{eq:firewall_thooft_1} and~\eqref{eq:firewall_thooft_2} entirely within the Hamiltonian formalism, which will provide a clearer path to quantization and will enable the resulting physics to be more easily interpretable. 

In this paper we use an entirely Hamiltonian framework for both the particle dynamics and gravitational effects by invoking the 
formalism developed by Arnowitt, Deser, and Misner (ADM)~\citep{arnowitt_2008}, which was subesquently specialized to the spherically symmetric problem by Kuchar~\citep{Kuchar_1994}. We use a model consisting of a background Schwarzschild black hole with two thin spherical shells of null matter centered at the singularity, one ingoing and one outgoing. This is the same model as considered in~\citep{dray_1985_2,Hajicek_2001_null_shells_1,Hajicek_2001_null_shells_2,Kouletsis_2001_null_shells_3}, with the added restriction that the two shells intersect outside all event horizons. Similar two-shell models have also been considered.~\citep{Bicak_2003,Jezierski_2003}  A further advantage of our derivation of the classical firewall transformation is that it requires no approximation; previous considerations required small shell energies and a collision near the event horizon of the black hole, albeit the models were for particles, not spherical shells.~\citep{'thooft_2016,'thooft_2016_2,'thooft_2018} As a result, our classical analogs of~\eqref{eq:firewall_thooft_1} and~\eqref{eq:firewall_thooft_2} will differ in some details from 't Hooft's.

The structure of the paper is as follows. In Section~\ref{sec:ADM_review}, we review the ADM formalism with spherical symmetry and include two spherical shells of null matter, one ingoing and one outgoing. In Section~\ref{sec:matching_conditions}, we characterize the discontinuities of the gravitational degrees of freedom at each shell, which include consistency conditions on the canonical variables that must apply at the event where the shells intersect. In Section~\ref{sec:kruskal_embedding}, we embed the ADM foliation coordinates in Kruskal-like coordinate systems in order to physically interpret the results of Sections~\ref{sec:ADM_review} and~\ref{sec:matching_conditions} as well as derive what is analogous to the firewall transformation in the ADM formalism.  In Section~\ref{sec:discussion}, we contextualize our results with the formalism of 't Hooft and discuss future paths to the quantization of our results.

\section{The ADM formalism with spherical symmetry and two shells}
\label{sec:ADM_review}
Here we review the ADM Hamiltonian formalism for general relativity with spherical symmetry and include two uniform shells of null matter. In summary, the ADM formalism foliates the spacetime with hypersurfaces of constant ``time'', allowing us to treat general relativity as an initial value problem, where we are given initial data on a given hypersurface, which we can evolve forward or backward in the given time coordinate. The ADM formalism, assuming spherical symmetry, admits the following decomposition of the metric:~\citep{louko_1998}
\begin{align}
    \rmd s^2&=g_{\mu\nu}\rmd x^\mu \rmd x^\nu\nonumber\\
    &=-N^2\rmd t^2+\Lambda^2(\rmd r+N^r \rmd t )^2 +R^2\rmd\Omega^2\nonumber\\
    &=-(N^2-\Lambda^2 N^r{}^2)\rmd t^2+2\Lambda^2 N^r \rmd t \rmd r + \Lambda^2 \rmd r^2+R^2 \rmd\Omega^2,\label{eq:foliation_metric}
\end{align}
where $\rmd s$ is a line element in the geometry, $N(t,r)$ is the lapse, $N^r(t,r)$ is the shift, $\Lambda(t,r)$ and $R(t,r)$ are the canonical variables of the metric, and $\rmd\Omega^2=\rmd\theta^2+\sin^2\theta \rmd\phi^2$. Because we have assumed spherical symmetry, the angular part of the metric only enters into the dynamics via $R(t,r)$, the circumferential radial coordinate. The lapse and shift are arbitrary continuous functions of $r$ and $t$ that fix the foliation and thus the coordinates $r$ and $t$. We assume that $(t,r,\theta,\phi)$ form a continuous coordinate system in the sense that $N$, $N^r$, $\Lambda$, and $R$ are all $C^0$ functions of $r$ and $t$. 

The ADM action (equivalent to the Einstein-Hilbert action) for general relativity with spherical symmetry is~\citep{Kuchar_1994}
\begin{align}
    S_G&=\int \rmd t \int \rmd r \Big(-\frac{1}{N}\Big(R\big(\dot{\Lambda}-(\Lambda N^r)'\big)(\dot{R}-R'N^r)+\tfrac{1}{2}\Lambda(\dot{R}-R'N^r)^2\Big)\nonumber\\
    &\hspace{10em}+N\Big(\frac{RR'\Lambda'}{\Lambda^2}-\frac{RR''}{\Lambda}-\frac{R'^2}{2\Lambda} + \frac{\Lambda}{2}\Big)\Big),
\end{align}
where a dot signifies a partial derivative with respect to $t$ and a prime signifies a partial derivative with respect to $r$. The conjugate momenta are~\citep{louko_1998}
\begin{align}
    P_\Lambda &= -\frac{R}{N}(\dot{R}-N^r R'),\\
    P_R &= -\frac{\Lambda}{N}(\dot{R}-N^r R')-\frac{R}{N}(\dot{\Lambda}-(N^r\Lambda)').
\end{align}
We construct our toy model by introducing a matter action representing two spherical shells, which move only radially:
\begin{align}
    S_M&=-m_\text{in}\int \rmd t \sqrt{N^2_\text{in}-\Lambda^2_\text{in}(\dot{\mathfrak{r}}_\text{in}+N^r_\text{in})^2}-m_\text{out}\int \rmd t \sqrt{N^2_\text{out}-\Lambda^2_\text{out}(\dot{\mathfrak{r}}_\text{out}+N^r_\text{out})^2},\label{eq:particle_action}
\end{align} 
where $\mathfrak{r}_{\text{in}/\text{out}}(t)$ is the canonical coordinate $r$ of the ingoing/outgoing shell's trajectory, respectively, and the subscripts ``in'' and ``out'' signify that the quantity is evaluated at $r=\mathfrak{r}_{\text{in}/\text{out}}(t)$, respectively. For the remainder of the paper, we will suppress the subscripts ``in'' and ``out'' when writing an equation that applies with both subscripts separately and unambiguously. The canonical momentum conjugate to each coordinate $\mathfrak{r}_{\text{in}/\text{out}}$ is
\begin{align}
\mathfrak{p}=\frac{m\Lambda^2(\dot{\mathfrak{r}}+N^r)}{\sqrt{N^2-\Lambda^2(\dot{\mathfrak{r}}+N^r)^2}}.
\end{align}
Solving directly for the velocity, we find
\begin{align}
    \dot{\mathfrak{r}}=\eta\frac{N/\Lambda}{\sqrt{1+\Lambda^2m^2/\mathfrak{p}^2}}-N^r\label{eq:rdot_m},
\end{align}
where $\eta_\text{in/out}=\operatorname{sign}(\mathfrak{p}_\text{in/out})$. The value of $\eta_\text{in/out}$ distinguishes an ingoing shell from an outgoing shell, so by definition we impose $\eta_\text{in}=-1$ and $\eta_\text{out}=1$. For the remainder of the paper we consider only the null limit $m_{\text{in/out}}\rightarrow 0$. Then
\begin{align}
    \dot{\mathfrak{r}}=\eta\frac{N}{\Lambda}-N^r.
\end{align}
Because $N$, $\Lambda$, and $N^r$ are all continuous, $\dot{\mathfrak{r}}_\text{in/out}$ is thus unambiguous along each shell. The full action written in terms of the Hamiltonian is then
\begin{align}
    S&=S_M+S_G\nonumber\\
    &=\int \rmd t\Big(\mathfrak{p}_\text{in} \dot{\mathfrak{r}}_\text{in}+\mathfrak{p}_\text{out}\dot{\mathfrak{r}}_\text{out}+\int dr(P_\Lambda\dot{\Lambda} + P_R \dot{R}-NH-N^r H_r)\Big),\label{eq:h_action}
\end{align}
where
\begin{align}
    H &= \frac{\Lambda P_\Lambda^2}{2R^2}-\frac{P_\Lambda P_R}{R} +\frac{RR''}{\Lambda} -\frac{RR'\Lambda'}{\Lambda^2}+\frac{R'{}^2}{2\Lambda}-\frac{\Lambda}{2}\nonumber\\
    &\hspace{10em}+\frac{\eta_\text{in} \mathfrak{p}_\text{in}}{\Lambda}\delta(r-\mathfrak{r}_\text{in})+\frac{\eta_\text{out} \mathfrak{p}_\text{out}}{\Lambda}\delta(r-\mathfrak{r}_\text{out}),\\
    H_r &= P_R R'-P'_\Lambda \Lambda - \mathfrak{p}_\text{in}\delta(r-\mathfrak{r}_\text{in})- \mathfrak{p}_\text{out}\delta(r-\mathfrak{r}_\text{out}),
\end{align}
where $\delta(r-\mathfrak{r}_\text{in/out})$ is a Dirac delta function in $r$. 

Hamilton's equations of motion are analogous to those found in~\citep{louko_1998}:
\begin{align}
    \dot{\Lambda}&=N\Big(\frac{\Lambda P_\Lambda}{R^2}-\frac{P_R}{R}\Big) + (N^r \Lambda)',\label{eq:Lambdadot}\\
    \dot{R} &= -\frac{NP_\Lambda}{R} + N^r R',\label{eq:Rdot}\\
    \dot{P}_\Lambda &= \frac{1}{2}N\Big( - \frac{P_\Lambda^2}{R^2}-\Big(\frac{R'}{\Lambda}\Big)^2+1+\frac{2\eta_\text{in} \mathfrak{p}_\text{in}}{\Lambda^2}\delta(r-\mathfrak{r}_\text{in})+\frac{2\eta_\text{out} \mathfrak{p}_\text{out}}{\Lambda^2}\delta(r-\mathfrak{r}_\text{out})\Big)\nonumber\\
    &\hspace{5em}- \frac{N' RR'}{\Lambda^2}+N^r P_\Lambda',\label{eq:PLambdadot}\\
    \dot{P}_R&=N\Big(\frac{\Lambda P_\Lambda^2}{R^3}-\frac{P_\Lambda P_R}{R^2}-\Big(\frac{R'}{\Lambda}\Big)'\Big)-\Big(\frac{N' R}{\Lambda}\Big)'+(N^rP_R)',\label{eq:PRdot}\\
    \dot{\mathfrak{r}}&=\eta \frac{N}{\Lambda}-N^r,\label{eq:rdot}\\
    \dot{\mathfrak{p}} &=-\mathfrak{p}\Big(\eta\frac{N}{\Lambda}-N^r\Big)'\Big|_{r=\mathfrak{r}}\label{eq:pdot}.
\end{align}
Equations~\eqref{eq:Lambdadot} through~\eqref{eq:pdot} along with the Hamiltonian constraints~\citep{louko_1998}
\begin{align}
    H&=0,\label{eq:h_constraint}\\
    H_r&=0\label{eq:h_constraint_r}
\end{align}
fully describe the dynamics of the model. In other words, for arbitrary $N$ and $N^r$, given initial data on a hypersurface of constant $t$ that satisfy the Hamiltonian constraints~\eqref{eq:h_constraint} and~\eqref{eq:h_constraint_r}, we can evolve the canonical variables to anywhere on the entire spacetime using~\eqref{eq:Lambdadot} through~\eqref{eq:pdot}.

\section{The equations of motion and constraints at the shells}
\label{sec:matching_conditions}

\begin{figure}
    \centering
    \includegraphics{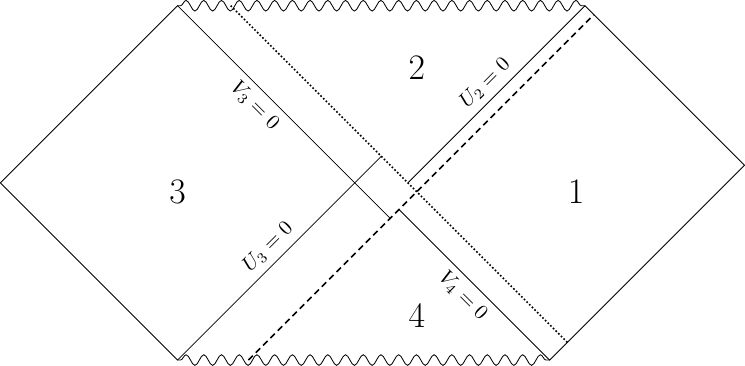}
    \caption{A Penrose diagram for a Schwarzschild black hole spacetime including two spherical shells of null matter, one ingoing and one outgoing. The ingoing shell is the dotted line and the outgoing shell is the dashed line. Lines of constant $U$ run from bottom left to top right and lines of constant $V$ run from bottom right to top left, both at an angle of 45 degrees. The diagram is partitioned into four labeled regions, separated by the two shells, which collide exterior to the event horizon. Note that region 1 lies entirely within the exterior of the black hole, while regions 2, 3, and 4 each contain part of the exterior and part of the interior.  Each of the four regions is installed with its own set of Kruskal coordinates $(U_i,V_i)$ and Schwarzschild mass $M_i$, and the visible coordinate axes are labeled in each region. Because the event horizon of the black hole moves when matter crosses it, the $U$ coordinate shifts when crossing the ingoing shell and the $V$ coordinate shifts when crossing the outgoing shell.}
    \label{fig:penrose_diagram}
\end{figure}

In the previous section, we derived the equations of motion for the canonical coordinates $\Lambda$, $R$, and $\mathfrak{r}_i$ and the canonical momenta $P_\Lambda$, $P_R$, and $\mathfrak{p}_i$. Off the spherical shells of null matter, the equations of motion are equivalent to the vacuum Einstein field equations. Because our toy model is spherically symmetric, the Schwarzschild spacetime will solve the metric everywhere off the shells. In Figure~\ref{fig:penrose_diagram}, we see a Penrose diagram of the entire spacetime, separated into four regions bounded by the two shells. However, we must take great care in how to relate the Schwarzschild spacetimes to each other at the boundaries formed by each shell.

We now turn to the equations of motion to determine what happens to the canonical variables as we cross the shells. By construction, $\Lambda$, $R$, and $\mathfrak{r}_{\text{in}/\text{out}}$ are all continuous across the shells. However,~\eqref{eq:PLambdadot} and~\eqref{eq:PRdot} show that $P_\Lambda$ and $P_R$ may be discontinuous across the shells, and~\eqref{eq:Lambdadot} and~\eqref{eq:Rdot} show that $\dot{R}$ and $\dot{\Lambda}$ will inherit any such discontinuity. Furthermore, introducing the notation
\begin{align}
    \dot{R}_\text{in/out}&=\partial_t(R)|_{r=\mathfrak{r}_\text{in/out}},
\end{align}
using the chain rule we write
\begin{align}
    \frac{\rmd}{\rmd t}(R(t,\mathfrak{r}(t)))=\dot{R}+R'\dot{\mathfrak{r}}.~\label{eq:dot_prime_shell}
\end{align}
Since $\frac{\rmd}{\rmd t}R(t,\mathfrak{r}_\text{in/out})$ and $\dot{\mathfrak{r}}_\text{in/out}$ are unambiguous along the shells by construction, $R'$ will inherit any discontinuity in $\dot{R}$, and similarly $\Lambda'$ will inherit any discontinuity from $\dot{\Lambda}$. 

In order to make the discontinuities explicit, we introduce the notation~\citep{louko_1998}
\begin{align}
    \Delta_\text{in/out} f\equiv \lim_{\epsilon\rightarrow 0+ }(f(t,\mathfrak{r}_\text{in/out}+\epsilon)-f(t,\mathfrak{r_\text{in/out}}-\epsilon)).\label{eq:delta_def}
\end{align}
for any function $f(t,r)$. In this notation, the delta contributions to $f'$ and $\dot{f}$ can be written as $(\Delta_\text{in/out} f)\delta(r-\mathfrak{r}_\text{in/out})$ and $-\dot{\mathfrak{r}}(\Delta_\text{in/out} f)\delta(r-\mathfrak{r}_\text{in/out})$, respectively. Now, if we integrate~\eqref{eq:h_constraint} and~\eqref{eq:h_constraint_r} from $r=\mathfrak{r_\text{in/out}}-\epsilon$ to $r=\mathfrak{r}_\text{in/out}+\epsilon$ at time $t$, we find
\begin{align}
    0 &= \frac{R}{\Lambda}\Delta R' +\frac{\eta \mathfrak{p}}{\Lambda},\label{eq:dH}\\
    0 &= -\Lambda\Delta P_\Lambda  - \mathfrak{p}.
\end{align}
We apply~\eqref{eq:delta_def} to~\eqref{eq:Lambdadot} and~\eqref{eq:Rdot} directly, and for~\eqref{eq:PLambdadot} and~\eqref{eq:PRdot} we again integrate from $r=\mathfrak{r_\text{in/out}}-\epsilon$ to $r=\mathfrak{r}_\text{in/out}+\epsilon$. With the aid of~\eqref{eq:dot_prime_shell}, its analog for $\Lambda$, and the fact that $\rmd\Lambda_\text{in/out}/\rmd t$ and $\rmd R_\text{in/out}/\rmd t$ are both unambiguous along the shells, we find
\begin{align}
    -(\eta\frac{N}{\Lambda}-N^r)\Delta\Lambda'&=N\Big(\frac{\Lambda \Delta P_\Lambda}{R^2}-\frac{\Delta P_R}{R}\Big) + \Lambda \Delta N^r{}'+N^r \Delta \Lambda',\label{eq:dLambdadot}\\
    -(\eta\frac{N}{\Lambda}-N^r)\Delta R' &= -\frac{N\Delta P_\Lambda}{R} + N^r \Delta R',\label{eq:dRdot}\\
    -(\eta\frac{N}{\Lambda}-N^r)\Delta P_\Lambda &= \frac{1}{2}N\frac{2\eta \mathfrak{p}}{\Lambda^2} +N^r \Delta P_\Lambda,\label{eq:dPLambdadot}\\
    -(\eta\frac{N}{\Lambda}-N^r)\Delta P_R&=-N \frac{\Delta R'}{\Lambda}-\frac{R\Delta N' }{\Lambda}+N^r\Delta P_R\label{eq:dPRdot}.
\end{align}
Of the six equations~\eqref{eq:dH} through~\eqref{eq:dPRdot}, only four are independent, which can be written
\begin{align}
    \Delta R' &= -\frac{\eta \mathfrak{p}}{R},\label{eq:dRprime}\\
    \Delta P_\Lambda &=   - \frac{\mathfrak{p}}{\Lambda}\label{eq:dplambda},\\
     \Delta\Lambda'&=\frac{\Lambda}{N}\Delta N'-\eta\frac{\Lambda^2}{N}\Delta N^r{}',\label{eq:dNrprime}\\
       \Delta P_R&=\eta\frac{R}{N}\Delta N'-\frac{ \mathfrak{p}}{R}.\label{eq:dNprime}
\end{align}
We can now see that the discontinuities in $\Lambda'$ and $P_R$ are arbitrary because they depend on the arbitrary choice of $N'$ and $N^r{}'$. In particular, there exists a unique set of choices $\Delta_\text{in/out} N'=\eta_\text{in/out}N\mathfrak{p}_\text{in/out}/R^2$ and $\Delta_\text{in/out} N^r{}'=\eta_\text{in/out} (\Delta_\text{in/out} N')/\Lambda$ such that both $\Lambda'$ and $P_R$ are continuous at both shells. However, $R'$ and $P_\Lambda$ remain unavoidably discontinuous at the shells as long as $\mathfrak{p}_\text{in/out}\neq 0$. 

The discontinuities described in~\eqref{eq:dRprime} through~\eqref{eq:dNprime} must be consistent when applied sequentially clockwise or counter-clockwise through all four regions of Figure~\ref{fig:penrose_diagram}. Despite this, one still may expect the discontinuity of a quantity across one shell to change when crossing the opposite shell. To check for this possiblity, we enforce the consistency of the discontinuities and make algebraic substitutions using~\eqref{eq:dRprime} through~\eqref{eq:dNprime} to see how the discontinuities on opposite sides of the shells compare. Doing so for $N'$, we find
\begin{align}
    0&=-(N'|_1-N'|_2)-(N'|_2-N'|_3)+(N'|_4-N'|_3)+(N'_1-N'|_4)\nonumber\\
    &=\frac{N}{R}(-\eta_\text{out}(P_R|_1-P_R|_2)-\eta_\text{in}(P_R|_2-P_R|_3)\nonumber\\
    &\hspace{12em}+\eta_\text{out}(P_R|_4-P_R|_3)+\eta_\text{in}(P_R|_1-P_R|_4))\nonumber\\
    &=2\frac{N}{R}(-P_R|_1+P_R|_2-P_R|_3+P_R|_4),~\label{eq:dd_1_original}
\end{align}
where all quantities are evaluated at the collision event and $|_i$ signifies that the quantity is evaluated in region $i$. At the risk of a minor abuse of notation, we can summarise the result of~\eqref{eq:dd_1_original} in the form
\begin{align}
    0&=\Delta_{\text{out}}(\Delta_\text{in}P_R),\label{eq:dd_1}
\end{align}
where all quantities are evaluated at the point of collision. Analogous calculations for the remaining variables show
\begin{align}
    0&=\Delta_{\text{out}}(\Delta_\text{in}N'),\label{eq:dd_2}\\
    0&=\Delta_{\text{out}}(\Delta_\text{in}N^r{}'),\label{eq:dd_3}\\
    0&=\Delta_{\text{out}}(\Delta_\text{in}\Lambda'),\label{eq:dd_4}\\
    0&=\Delta_{\text{out}}(\Delta_\text{in}P_\Lambda),\label{eq:dd_5}\\
    0&=\Delta_{\text{out}}(\Delta_\text{in}R').\label{eq:dd_6}
\end{align}
It is also straightforward to show that the operations $\Delta_{\text{out}}\Delta_\text{in}$ and $\Delta_{\text{in}}\Delta_\text{out}$ are equivalent. Thus, the discontinuities across one shell are the same on both sides of the opposite shell at the point of collision.

Now we will discuss the continuity of the shell momenta. Equation~\eqref{eq:rdot} is continuous across the shells by construction, but for~\eqref{eq:pdot} we calculate
\begin{align}
     \Delta_\text{in}\dot{\mathfrak{p}}_\text{in}&= -\mathfrak{p}_\text{in}\Delta_\text{in}\Big(\eta_\text{in}\frac{N}{\Lambda}-N^r\Big)'\Big|_{r=\mathfrak{r}_\text{in}}\nonumber\\
     &=-\mathfrak{p}_\text{in}\Big(\eta_\text{in} \frac{1}{\Lambda}\Delta_\text{in}N'-\eta_\text{in} \frac{N}{\Lambda^2}\Delta_\text{in}\Lambda'-\Delta_\text{in}N^r{}'\Big)\Big|_{r=\mathfrak{r}_\text{in}}\nonumber\\
     &=0,\\
    \Delta_\text{out}\dot{\mathfrak{p}}_\text{out}&=0,
\end{align}
where we've applied~\eqref{eq:dNrprime}. Thus, along its own shell, $\dot{\mathfrak{p}}_\text{in/out}$ is unambiguous along the shell (i.e. it is the same on both sides of its own shell). A significant feature of the toy model is that the shells move in opposite directions and will thus collide at some time $t=t_0$, where $\mathfrak{r}_\text{in}(t_0)=\mathfrak{r}_\text{out}(t_0)$. At the collision event $(t_0,\mathfrak{r}_\text{in/out}(r_0))$, we can see how $\dot{\mathfrak{p}}_\text{in/out}$ changes after crossing the other shell:
\begin{align}
     \Delta_\text{out}\dot{\mathfrak{p}}_\text{in}(t_0)&=-\mathfrak{p}_\text{in}\Big(\eta_\text{in} \frac{1}{\Lambda}\Delta_\text{out}N'-\eta_\text{in} \frac{N}{\Lambda^2}\Delta_\text{out}\Lambda'-\Delta_\text{out}N^r{}'\Big)\Big|_{r=\mathfrak{r}_\text{in/out}(t_0)}\nonumber\\
     &=-\mathfrak{p}_\text{in}(\eta_\text{in}\eta_\text{out}-1)\Delta_\text{out} N^r{}'\Big|_{r=\mathfrak{r}_\text{in/out}(t_0)}\nonumber\\
     &=2\mathfrak{p}_\text{in}\Delta_\text{out}N^r{}'\Big|_{r=\mathfrak{r}_\text{in/out}(t_0)},\label{eq:dpin_dot}\\
     \Delta_\text{in}\dot{\mathfrak{p}}_\text{out}(t_0)&=2\mathfrak{p}_\text{out}\Delta_\text{in}N^r{}'\Big|_{r=\mathfrak{r}_\text{in/out}(t_0)}.\label{eq:dpout_dot}
\end{align}
Thus, $\dot{\mathfrak{p}}_\text{in/out}$ may be discontinuous when crossing the opposite shell, though $\mathfrak{p}_\text{in/out}$ remains continuous.

Equations~\eqref{eq:dRprime} through~\eqref{eq:dNprime} along with ~\eqref{eq:rdot} and~\eqref{eq:pdot} fully characterize the dynamics of the spacetime along and across the two shells. Importantly, they directly show the effect of the shell momenta $\mathfrak{p}_\text{in/out}$ on the gravitational degrees of freedom.

\section{An embedding in generalized Kruskal coordinates}
\label{sec:kruskal_embedding}
Up until now, we have only used the arbitrary, continuous foliation coordinates $(t,r)$ to describe the spacetime and shell variables, but in order to extract meaningful physical information from the results above, we must relate our results to physically interpretable coordinate systems. For example, the Schwarzschild metric is given by
\begin{align}
    \rmd s^2=-F_i \rmd T_i^2+F_i^{-1}\rmd R^2+R^2\rmd\Omega^2,\label{eq:schwarzschild}
\end{align}
where $T_i$ is the Schwarzschild time and
\begin{align}
    F_i=1-\frac{2M_i}{R}.
\end{align}
Writing $T=T(t,r)$ and $R=(t,r)$, one can show that for appropriate choices of $N$ and $N^r$, the Schwarzschild metric solves~\eqref{eq:h_constraint} through~\eqref{eq:h_constraint_r} off the shells.~\citep{Kuchar_1994} In other words, we can install Schwarzschild coordinates $(T_i,R)$ in each region $i\in\{1,2,3,4\}$ of Figure~\ref{fig:penrose_diagram} separately, with different Schwarzschild masses $M_i$. Note that the Schwarzschild radial coordinate $R$ is the same in each region because it is the circumferential radial coordinate, and thus is continuous across the shells, but the Schwarzschild time $T_i$ is not the same in the different regions. For the remainder of the section, we generally suppress the subscripts $i$ in all equations defined in each region $i$ of Figure~\ref{fig:penrose_diagram} separately. A wealth of information can be extracted by embedding the foliation coordinates $(t,r)$ in various coordinate systems. In order to derive the firewall transformation, we investigate two Kruskal-like coordinate systems.

\subsection{Generalized Kruskal Coordinates}
In order to derive the firewall transformation, we are interested in a Kruskal-like embedding of the foliation, similar to the formulation in~\citep{Hajicek_2001}. The traditional Kruskal coordinates $(\mathcal{U}_i, \mathcal{V}_i)$ we define separately in each region:
\begin{align}
    \rmd s^2=\frac{32 M^3}{R}\rme^{-R/2M} \rmd\mathcal{U} \rmd\mathcal{V} + R^2 \rmd\Omega^2,
\end{align}
where
\begin{align}
     \mathcal{U} \mathcal{V}&=\Big(\frac{R}{2M}-1\Big)\rme^{R/2M},\label{eq:uv_r}\\
    \frac{ \mathcal{V}}{ \mathcal{U}}&=\rme^{ T/2M}\label{eq:uv_t}
\end{align}
where $R$ is to be understood as a function of $\mathcal{U}_i$ and $\mathcal{V}_i$ implicitly given by~\eqref{eq:uv_r}. The definition of Kruskal-like coordinates admits an overall multiplicative factor for each coordinate. Though not needed in this section, we nevertheless introduce such multiplicative factors here for use in Section~\ref{sec:adm_firewall}.  Thus, we define generalized Kruskal coordinates $(U_i,V_i)$:
\begin{align}
    U&\equiv \beta \mathcal{U},\\
    V&\equiv \gamma \mathcal{V},
\end{align}
where the $\beta_i$ and $\gamma_i$ constitute eight constants to be determined later. The metric for the generalized Kruskal coordinates is
\begin{align}
    \rmd s^2=\frac{1}{\beta \gamma}\frac{32 M^3}{ R}\rme^{-R/2M} \rmd U \rmd V + R^2 \rmd\Omega^2,\label{eq:generalized_kruskal}
\end{align}
where
\begin{align}
     U V&=\beta \gamma\Big(\frac{R}{2M}-1\Big)\rme^{R/2M},\label{eq:uvt_r}\\
    \frac{ V}{ U}&=\frac{\gamma}{\beta}\rme^{ T/2M}\label{eq:uvt_t}.
\end{align}

We can embed the foliation coordinates $(t,r)$ in the generalized Kruskal-Szekeres metric by assuming a coordinate transformation exists of the form $U_i(t,r)$ and $V_i(t,r)$. We find
\begin{align}
    \rmd s^2&=\frac{1}{\beta \gamma}\frac{32 M^3}{ R}\rme^{-R/2M} (\dot{U}\rmd t+U' dr)(\dot{V} \rmd t +V'\rmd r)+R^2\rmd\Omega^2\nonumber\\
    &=\frac{1}{\beta \gamma}\frac{32 M^3}{ R}\rme^{-R/2M} \Big(\dot{U}\dot{V}\rmd t^2+(\dot{U} V'+ U'\dot{V})\rmd t\rmd r+U'V'\rmd r^2\Big)+R^2\rmd\Omega^2.\label{eq:g_rt_UV}
\end{align}
Comparing with~\eqref{eq:foliation_metric} yields
\begin{align}
    N^2&=\frac{1}{4}\frac{1}{\beta \gamma}\frac{32 M^3}{ R}\rme^{-R/2M} U'V'\Big(\frac{\dot{U}}{U'}-\frac{\dot{V}}{V'}\Big)^2\nonumber\\
    &=4M^2F\frac{U'}{U}\frac{V'}{V}\Big(\frac{\dot{U}}{U'}-\frac{\dot{V}}{V'}\Big)^2,\label{eq:n_kruskal}\\
    N^r&=\frac{1}{2}\Big(\frac{\dot{U}}{U'}+\frac{\dot{V}}{V'}\Big),\label{eq:nr_kruskal}\\
    \Lambda^2& = \frac{1}{\beta \gamma}\frac{32 M^3}{ R}\rme^{-R/2M} U'V',\nonumber\\
    &=16M^2F\frac{U'}{U}\frac{V'}{V}\label{eq:lambda_kruskal}
\end{align}
applied in each region $i$ separately, and we see that the $\beta_i$ and $\gamma_i$ have disappeared completely. Because $N^r$, $\Lambda$, and $N$ are everywhere continuous, we also have information on how the Kruskal coordinates change across the shells in terms of the foliation coordinates. In the notation of the previous section, at the shells~\eqref{eq:n_kruskal},~\eqref{eq:nr_kruskal}, and~\eqref{eq:lambda_kruskal} can be rearranged to find
\begin{align}
    0&=\Delta\Big(\frac{\dot{U}}{U'}+\frac{\dot{V}}{V'}\Big),\\
    0&=\Delta\Big(\frac{\dot{U}\dot{V}}{U'V'}\Big),\\
    0&=\Delta\Big(M^2F\frac{U'}{U}\frac{V'}{V}\Big)\label{eq:delta_lambda}.
\end{align}
It is also straightforward to calculate $\mathfrak{r}_\text{in/out}$ directly:
\begin{align}
    \dot{\mathfrak{r}}_\text{in}&=-\frac{N_\text{in}}{\Lambda_\text{in}}-N^r_\text{in}=-\frac{\dot{V}_\text{i,in}}{V'_\text{i,in}}\label{eq:rdot_in_k}\\
    \dot{\mathfrak{r}}_\text{out}&=\frac{N_\text{out}}{\Lambda_\text{out}}-N^r_\text{out}=-\frac{\dot{U}_\text{i,out}}{U'_\text{i,out}}.
\end{align}
Using the chain rule, we can also write
\begin{align}
    \frac{d}{\rmd t}(V_\text{in})&=\dot{V}_\text{in}+V_{\text{in}}'\dot{\mathfrak{r}}_\text{in},\label{eq:Vd_chain}\\
    \frac{d}{\rmd t}(U_\text{out})&=\dot{U}_\text{out}+U_{\text{out}}'\dot{\mathfrak{r}}_\text{out}.\label{eq:Ud_chain}
\end{align}
Finally, combining~\eqref{eq:rdot_in_k} through~\eqref{eq:Ud_chain} we find
\begin{align}
    \frac{\rmd}{\rmd t}(V_\text{in})&=0,\label{eq:v_const}\\
    \frac{\rmd}{\rmd t}(U_\text{out})&=0.\label{eq:u_const}
\end{align}
We have now proven that the ingoing shell moves along constant $V_i$ and the outgoing shell moves along constant $U_i$ in each region.

\subsection{Eddington-Finkelstein-like coordinates}
We also find useful an embedding in Eddington-Finkelstein-like coordinates, where instead of the Eddington-Finkelstein coordinates $u=t-r_*$ and $v=t+r_*$ (where $r_*$ is the tortoise coordinate), we use the generalized Kruskal coordinates $U$ and $V$. By writing $U_i=U_i(V_i,R)$ or $V_i=(U_i,R)$ via~\eqref{eq:uvt_r}, we can find the metric in either $(U_i,R)$ or $(V_i,R)$ coordinates from~\eqref{eq:generalized_kruskal}:
\begin{align}
    \rmd s^2&=-F\Big(\frac{4M}{U}\Big)^2\rmd U^2+2\frac{4M}{U}\rmd U\rmd R+R^2\rmd\Omega^2,\\
    \rmd s^2&=-F\Big(\frac{4M}{V}\Big)^2\rmd V^2+2\frac{4M}{V}\rmd V\rmd R+R^2\rmd\Omega^2.
\end{align}
Assuming an embedding $U_i(t,r)$ and $V_i(t,r)$ separately in each region, we find
\begin{align}
    \rmd s^2&=-F\Big(\frac{4M}{U}\Big)^2(\dot{U}\rmd t+U'\rmd r)^2+2\frac{4M}{V}(\dot{U}\rmd t+U'\rmd r)(\dot{R}\rmd t+R'\rmd r)+R^2\rmd\Omega^2\nonumber\\
    &=-\frac{4M}{U}\Big(\frac{4M}{U}F\dot{U}^2-2\dot{U}\dot{R}\Big)\rmd t^2+2\frac{4M}{U}\Big(-\frac{4M}{U}F\dot{U}U'+U'\dot{R}+\dot{U}R'\Big)\rmd t\rmd r\nonumber\\
    &\hspace{10em}+\frac{4M}{U}\Big(-\frac{4M}{U}FU'{}^2+2U'R'\Big)\rmd r^2+R^2 \rmd\Omega^2,\label{eq:g_rt_UR}\\
    ds^2&=-\frac{4M}{V}\Big(\frac{4M}{V}F\dot{V}^2-2\dot{V}\dot{R}\Big)\rmd t^2+2\frac{4M}{V}\Big(-\frac{4M}{V}F\dot{V}V'+V'\dot{R}+\dot{V}R'\Big)\rmd t\rmd r\nonumber \\
    &\hspace{10em}+\frac{4M}{V}\Big(-\frac{4M}{V}FV'{}^2+2V'R'\Big)\rmd r^2+R^2 \rmd\Omega^2.\label{eq:g_rt_VR}
\end{align}
Comparing with~\eqref{eq:foliation_metric} yields
\begin{align}
    N^2&=2M\frac{U'}{U}\frac{(\dot{R} -R' \dot{U}/U')^2}{(R' -2MF (U'/U))}\\
    &=2M\frac{V'}{V}\frac{(\dot{R} -R' \dot{V}/V')^2}{(R' -2MF (V'/V))},\\
    \Lambda^2&= 8M\frac{U'}{U}\Big(R' -2M F  \frac{U'}{U}\Big)\\
    &= 8M\frac{V'}{V}\Big(R' -2M F  \frac{V'}{V}\Big),\\
    N^r&= \frac{1}{2}\frac{\dot{R} +R'  \dot{U}/U'-4M F  \dot{U} /U}{R'  -2 MF U'/U}\\
    &= \frac{1}{2}\frac{\dot{R} +R'  \dot{V}/V'-4M F  \dot{V} /V}{R'  -2 MF V'/V}.
\end{align}
We can see the new information we've gained in a compact form from the following combinations of the metric components and application of the chain rule:
\begin{align}
    \dot{\mathfrak{r}}g_{rr,\text{in}}+g_{tr,\text{in}}&=\frac{4M}{V_\text{in}}\Big(-\frac{4M}{V}\dot{\mathfrak{r}}_\text{in}FV'{}^2+2\dot{\mathfrak{r}}_\text{in}V'R'-\frac{4M}{V}F\dot{V}V'+V'\dot{R}+\dot{V}R'\Big)\Big|_{r=\mathfrak{r}_\text{in}}\nonumber\\
    &=4M\frac{V'_\text{in}}{V_\text{in}}\frac{\rmd}{\rmd t}R(t,\mathfrak{r}_\text{in}),\\
    \dot{\mathfrak{r}}g_{rr,\text{out}}+g_{tr,\text{out}}&=\frac{4M}{U_\text{out}}\Big(-\frac{4M}{U}\dot{\mathfrak{r}}FU'{}^2+2\dot{\mathfrak{r}}U'R'-\frac{4M}{U}F\dot{U}U'+U'\dot{R}+\dot{U}R'\Big)\Big|_{r=\mathfrak{r}_\text{out}}\nonumber\\
    &=4M\frac{U'_\text{out}}{U_\text{out}}\frac{\rmd}{\rmd t}R(t,\mathfrak{r}_\text{out}),
\end{align}
where we have used the chain rule,~\eqref{eq:v_const}, and~\eqref{eq:u_const}. Because $g_{rr}$, $g_{tr}$, and $\frac{\rmd}{\rmd t}R(t,\mathfrak{r}_\text{in/out})$ are all continuous across the shells, we have two new equations relating the Kruskal coordinates across the shells:
\begin{align}
    0&=\Delta_\text{in}\Big(M\frac{V'}{V}\Big),\label{eq:delta_vp}\\
    0&=\Delta_\text{out}\Big(M\frac{U'}{U}\Big).\label{eq:delta_up}
\end{align}
We can apply~\eqref{eq:delta_vp} and~\eqref{eq:delta_up} directly to $g_{rr}$ as found in~\eqref{eq:g_rt_VR} and~\eqref{eq:g_rt_UR}, respectively, to find the following conditions at the shells:
\begin{align}
    \Delta_\text{in} R'&=-\frac{4M}{R}\frac{\hat{V}'}{\hat{V}}\Delta_\text{in} M,\label{eq:rp_in}\\
    \Delta_\text{out} R'&=-\frac{4M}{R} \frac{\hat{U}'}{\hat{U}}\Delta_\text{out} M\label{eq:rp_out}
\end{align}

\subsection{Shell Momenta in terms of Schwarzschild masses}
Here we combine the results from the previous sections to relate the shell momenta $\mathfrak{p}_\text{in/out}$ to physical quantities. Combining~\eqref{eq:rp_in} and~\eqref{eq:rp_out} with~\eqref{eq:dRprime}, we find
\begin{align}
    \mathfrak{p}_\text{in}&=\eta_\text{in}\frac{4MV'_\text{in}}{V_\text{in}}\Delta_\text{in} M,\label{eq:pin_to_m}\\
    \mathfrak{p}_\text{out}&=\eta_\text{out}\frac{4MU'_\text{out}}{U_\text{out}}\Delta_\text{out} M.\label{eq:pout_to_m}
\end{align}
We can now calculate the canonical momenta directly from the energies of the shells, $\Delta_\text{in/out} M$, the Kruskal coordinates, and their relation to the foliation coordinates $(t,r)$. Note that the foliation remains completely arbitrary, so the canonical pair ($\mathfrak{r},\mathfrak{p}$) can refer to any spacetime coordinate for the shells and its conjugate momentum with an appropriate choice of $N$ and $N^r$.

\subsection{The Dray-'t Hooft-Redmount formula}
We now combine the results of this section in order to derive a constraint between the four masses $M_i$ and the radius $R$ of collision between the two shells. Firstly,~\eqref{eq:delta_vp} and~\eqref{eq:delta_up} give information on how $V'$ changes across the ingoing shell and how $U'$ changes across the outgoing shell. We can combine~\eqref{eq:delta_vp} and~\eqref{eq:delta_up} with~\eqref{eq:delta_lambda} in order to find how $U'$ and $V'$ change across the opposite shell:
\begin{align}
    0&=\Delta_\text{out}\Big(MF\frac{V'}{V}\Big),\\
    0&=\Delta_\text{in}\Big(MF\frac{U'}{U}\Big).\label{eq:delta_up_2}
\end{align}
Because one shell is ingoing and one is outgoing, the shells will collide and pass through each other at some time $t=t_0$, where $\mathfrak{r}_\text{in}(t_0)=\mathfrak{r}_\text{out}(t_0)$. We can then define the radius of collision $R_0\equiv R(t_0,\mathfrak{r}_\text{in/out}(t_0))$. We can now apply~\eqref{eq:delta_vp} through~\eqref{eq:delta_up_2} at the intersection point. For example, let us be given $V'_4(t_0,\mathfrak{r}_\text{in/out}(t_0))$. There are two possible paths to find $V'_2(t_0,\mathfrak{r}_\text{in/out}(t_0))$, one passing through region 1 and one passing through region 3. Explicitly,
\begin{align}
    V'_4(t_0,\mathfrak{r}_\text{in/out})(t_0)&=\frac{M_1}{M_4}\frac{V_4}{V_1}\Big(\frac{1-2M_1/R_0}{1-2M_4/R_0}\Big)V'_1(t_0,\mathfrak{r}_\text{in/out}(t_0))\nonumber\\
    &=\frac{M_2}{M_4}\frac{V_4}{V_2}\Big(\frac{1-2M_1/R_0}{1-2M_4/R_0}\Big)V'_2(t_0,\mathfrak{r}_\text{in/out}(t_0).\\
    V'_4(t_0,\mathfrak{r}_\text{in/out})(t_0)&=\frac{M_3}{M_4}\frac{V_4}{V_3}V'_3(t_0,\mathfrak{r}_\text{in/out}(t_0))\nonumber\\
    &=\frac{M_2}{M_4}\frac{V_4}{V_2}\Big(\frac{1-2M_2/R_0}{1-2M_3/R_0}\Big)V'_2(t_0,\mathfrak{r}_\text{in/out}(t_0).
\end{align}
If our coordinates are to be consistent at the point of intersection then we must have
\begin{align}
    \Big(1-\frac{2M_1}{R_0}\Big)\Big( 1-\frac{2M_3}{R_0}\Big)=\Big(1-\frac{2M_2}{R_0}\Big)\Big( 1-\frac{2M_4}{R_0}\Big).\label{eq:mass_cond}
\end{align}
Equation~\eqref{eq:mass_cond} is the Dray-'t Hooft-Redmount formula~\citep{dray_1985,Redmount_1985}, and is a result which is entirely independent of the choice of foliation and contains only physically measurable quantities.

We will now present several physical interpretations of the formula. Let us define the total energy in each shell before and after the collision by the difference in Schwarzschild masses along that shell:
\begin{align}
    E_\text{in}&=M_1-M_4,\\
    E_\text{out}&=M_4-M_3,\\
    \tilde{E}_\text{in}&=M_2-M_3,\\
    \tilde{E}_\text{out}&=M_1-M_2.\label{eq:eoutprime}
\end{align}
The four energies are not independent, since by construction
\begin{align}
    E_\text{in}+E_\text{out}=\tilde{E}_\text{in}+\tilde{E}_\text{out}.
\end{align}
Therefore, we may algebraically eliminate three masses in~\eqref{eq:mass_cond} in favor of three energies. For example, the following two equations are both equivalent to the Dray-'t Hooft-Redmount formula
\begin{align}
    \tilde{E}_\text{in}&=E_\text{in}+\frac{E_\text{in}E_\text{out}}{R_0-2M_4},\\
    \tilde{E}_\text{out}&=E_\text{out}-\frac{E_\text{in}E_\text{out}}{R_0-2M_4}.
\end{align}
Therefore, we can interpret the Dray-'t Hooft-Redmount formula as describing an exchange of energy equal to $E_\text{in}E_\text{out}/(R_0-2M_4)$ between the shells. Other useful rearrangements of the Dray-'t Hooft-Redmount formula include
\begin{align}
    \frac{E_\text{in}}{1-2M_4/R_0}&=\frac{\tilde{E}_\text{in}}{1-2M_3/R_0},\label{eq:mass_cond_momenta_1}\\
    \frac{E_\text{in}}{1-2M_1/R_0}&=\frac{\tilde{E}_\text{in}}{1-2M_2/R_0},\\
    \frac{E_\text{out}}{1-2M_4/R_0}&=\frac{\tilde{E}_\text{out}}{1-2M_1/R_0},\label{eq:mass_cond_momenta_3}\\
    \frac{E_\text{out}}{1-2M_3/R_0}&=\frac{\tilde{E}_\text{out}}{1-2M_2/R_0}.\label{eq:mass_cond_momenta_4}
\end{align}
In order to interpret~\eqref{eq:mass_cond_momenta_1} through~\eqref{eq:mass_cond_momenta_4}, we note that in a foliation tailored to the Schwarzschild coordinates in one region $t=T_i$ and $r=R$, the shell momenta become $\mathfrak{p}=\eta \Delta M/F$, according to~\eqref{eq:pin_to_m} and~\eqref{eq:pout_to_m}. Therefore, in Schwarzschild coordinates $(T_i,R)$, the momentum $\mathfrak{p}$ for a shell is different on both sides of that shell due to the change in Schwarzschild mass, but on a given side of the shell, the momentum remains the same before and after the collision. In other words, no radial momentum is transferred during the collision. Note, however, that we cannot enforce the foliation $t=T_i$ and $r=R$ in all regions $i$ simultaneously, as that would result in discontinuities in the metric functions $N$, $N^r$, and $\Lambda$.

\section{The ADM Analog of the Firewall Transformation}
\label{sec:adm_firewall}
Within the ADM framework we have developed thus far, we can finally find the shifts in the Kruskal coordinates of the shells, which form the basis for the firewall transformation. We showed in~\eqref{eq:v_const} and~\eqref{eq:u_const} that the ingoing shell moves along constant $V_i$ in all four regions, and the outgoing shell moves along constant $U_i$ in all four regions. We expect a shift in these constant Kruskal coordinates when the shells collide. In order for this shift to be consistent, we must enforce that along each shell before and after the collision, the constant Kruskal coordinate for that shell is the same on both sides of that shell. Explicitly, we enforce
\begin{align}
    V_{1,\text{in}}&=V_{4,\text{in}},\label{eq:consistency_14}\\
    V_{2,\text{in}}&=V_{3,\text{in}},\\
    U_{1,\text{out}}&=U_{2,\text{out}},\\
    U_{3,\text{out}}&=U_{4,\text{out}}.\label{eq:consistency_34}
\end{align}
We must ensure that~\eqref{eq:consistency_14} through~\eqref{eq:consistency_34} are consistent with the Dray-'t Hooft-Redmount formula. Equations~\eqref{eq:mass_cond} and~\eqref{eq:uvt_r} applied at the point of collision together with~\eqref{eq:consistency_14} through~\eqref{eq:consistency_34} imply
\begin{align}
    \beta_1\gamma_1\beta_3\gamma_3\frac{R_0^2}{4M_1 M_3}\exp\Big(\frac{R_0}{2M_1}+\frac{R_0}{2M_3}\Big)&=\beta_2\gamma_2\beta_4\gamma_4\frac{R_0^2}{4M_2M_4}\exp\Big(\frac{R_0}{2M_2}+\frac{R_0}{2M_4}\Big).\label{eq:mass_cond_gammabeta}
\end{align}
Therefore, we must choose the $\beta_i$ and $\gamma_i$ such that they satisfy~\eqref{eq:mass_cond_gammabeta}. To avoid mixing information between different regions, to ensure that $U_i V_i$ does not depend on $T_0$, and to ensure that $\lim_{R_0\rightarrow 2M_i} \beta_i \gamma_i \neq 0$ we make the choice
\begin{align}
    \beta_i \gamma_i&=\frac{2M_i}{R_0} \rme^{-R_0/2M_i}
\end{align}
in each region separately. The remaining freedom in $\beta_i$ and $\gamma_i$ is given by $\gamma_i/\beta_i$, as in~\eqref{eq:uvt_t}. We interpret this freedom as a shift in the Schwarzschild time in each region, defining four new constants
\begin{align}
    \tau_i\equiv 2M_i\ln\frac{\gamma_i}{\beta_i}
\end{align}
in each region separately. Now~\eqref{eq:uvt_r} and~\eqref{eq:uvt_t} are fully specified. Evaluating at the collision, we find
\begin{align}
    U_\text{out}V_\text{in}&= 1-\frac{2M}{R_0},\label{eq:uv_collision}\\
    \frac{V_\text{in}}{U_\text{out}}&= \rme^{(T_0-\tau)/2M},\label{eq:vu_collision}
\end{align}
in each region separately, where $T_{0,i}\equiv T_i(t_0,\mathfrak{r}(t_0))$. Therefore, the $\tau_i$ can be chosen to fix the time of collision in each region.  However, not all of these choices are independent because~\eqref{eq:consistency_14} through~\eqref{eq:consistency_34} now show explicitly how the Schwarzschild time changes across the shells:
\begin{align}
    \Big(1-\frac{2M_1}{R_0}\Big)^{1/2}\rme^{(T_{0,1}-\tau_1)/4M_1}&=\Big(1-\frac{2M_4}{R_0}\Big)^{1/2}\rme^{(T_{0,4}-\tau_4)/4M_4},\label{eq:T1_T4}\\
    \Big(1-\frac{2M_2}{R_0}\Big)^{1/2}\rme^{(T_{0,2}-\tau_2)/4M_2}&=\Big(1-\frac{2M_3}{R_0}\Big)^{1/2}\rme^{(T_{0,3}-\tau_3)/4M_3},\\
    \Big(1-\frac{2M_1}{R_0}\Big)^{1/2}\rme^{-(T_{0,1}-\tau_1)/4M_1}&=\Big(1-\frac{2M_2}{R_0}\Big)^{1/2}\rme^{-(T_{0,2}-\tau_2)/4M_2},\\
    \Big(1-\frac{2M_3}{R_0}\Big)^{1/2}\rme^{-(T_{0,3}-\tau_3)/4M_3}&=\Big(1-\frac{2M_4}{R_0}\Big)^{1/2}\rme^{-(T_{0,4}-\tau_4)/4M_4}.\label{eq:T3_T4}
\end{align}
Only three of the four equations are independent, as multiplying all four together gives the Dray-'t Hooft-Redmount formula. Thus, once $T_{0,i}-\tau_i$ is determined in one region $T_{0,i}-\tau_i$ can be calculated in all other regions from $R_0$ and the masses. If necessary, we have the freedom to choose the $\tau_i$ such that the Schwarzschild time of the collision $T_{0,i}$ is the same in all four regions. Even with this choice, the Schwarzschild times $T_i$ will still differ elsewhere along either side of the shells.

Let us pause and consider the initial value problem of the toy model. Suppose we are given physical data on a spacelike slice before the collision of the shells. For example, a sufficient set of data includes the mass in region 4, $M_4$, the two shell energies before the collision, $E_\text{in}$ and $E_\text{out}$, and the modified Kruskal coordinates of the shells, $U_\text{out,4}$ and $V_\text{in,4}$. We can then calculate all physical quantities in the spacetime. The radius of the collision, $R_0$, and the (shifted by $\tau$) Schwarzschild time of the collision in region 4, $T_{0,4}-\tau_4$, can be calculated via~\eqref{eq:uv_collision} and~\eqref{eq:vu_collision}, respectively. The radius of the collision, $R_0$, is the same in each region, the Schwarzschild masses and energies of the shells in all other regions are given by~\eqref{eq:mass_cond} through~\eqref{eq:eoutprime}, and the Schwarzschild times of the collisions in the other three regions are given by~\eqref{eq:T1_T4} through~\eqref{eq:T3_T4}. One can then calculate the $U_i$ and $V_i$ via~\eqref{eq:uv_collision} and~\eqref{eq:vu_collision}, which will automatically satisfy~\eqref{eq:consistency_14} through~\eqref{eq:consistency_34}. Thus, we have a complete picture of the spacetime given the five quantities $M_4$, $E_\text{in}$, $E_\text{out}$,  $U_\text{out,4}$, and $V_\text{in,4}$.

In order to present results resembling 't Hooft's firewall transformation, we now calculate the shifts in the shells' constant Kruskal coordinates before and after the collision in all regions:
\begin{align}
    V_{1,\text{in}}-V_{2,\text{in}}&=\frac{1}{U_{1,\text{out}}}\Big(1-\frac{2M_1}{R_0}\Big)-\frac{1}{U_{2,\text{out}}}\Big(1-\frac{2M_2}{R_0}\Big)\nonumber\\
    &=-\frac{2}{R_0U_{1,\text{out}}}\tilde{E}_\text{out},\\
    V_{4,\text{in}}-V_{3,\text{in}}&=-\frac{2}{R_0U_{4,\text{out}}}E_\text{out},\\
    U_{1,\text{out}}-U_{4,\text{out}}&=-\frac{2}{R_0V_{1,\text{out}}}E_\text{in},\\
    U_{2,\text{out}}-U_{3,\text{out}}&=-\frac{2}{R_0V_{2,\text{out}}}\tilde{E}_\text{in}.
\end{align}
The shift in a shell's constant Kruskal coordinate is the same on either side of the shell. For example,
\begin{align}
    V_{1,\text{in}}-V_{2,\text{in}}&=-\frac{2}{R_0U_{1,\text{out}}}\tilde{E}_\text{out}\nonumber\\
    &=-\frac{2V_1}{R_0}\frac{\tilde{E}_\text{out}}{(1-2M_1/R_0)},\nonumber\\
    &=-\frac{2V_4}{R_0}\frac{E_\text{out}}{(1-2M_4/R_0)}\nonumber\\
    &=-\frac{2}{R_0U_{4,\text{out}}}E_\text{out}\nonumber\\
    &=V_{4,\text{in}}-V_{3,\text{in}},
\end{align}
where we have employed the Dray-'t Hooft-Redmount formula in the form of~\eqref{eq:mass_cond_momenta_3}. Along with~\eqref{eq:pin_to_m} and~\eqref{eq:pout_to_m}, we can compactly write the shifts in the shells' constant Kruskal coordinates in terms of the shells' canonical momenta:
\begin{align}
    V_{1,\text{in}}-V_{2,\text{in}}&=-\frac{1}{2M_1 R_0 }\frac{\eta_\text{out} \mathfrak{p}_\text{out}}{U_{1,\text{out}}'},\label{eq:firewall_transformation_1}\\
    V_{4,\text{in}}-V_{3,\text{in}}&=-\frac{1}{2M_4 R_0 }\frac{\eta_\text{out} \mathfrak{p}_\text{out}}{U_{4,\text{out}}'},\label{eq:firewall_transformation_2}\\
    U_{1,\text{out}}-U_{4,\text{out}}&=-\frac{1}{2M_1 R_0 }\frac{\eta_\text{in} \mathfrak{p}_\text{in}}{V_{1,\text{in}}'},\label{eq:firewall_transformation_3}\\
    U_{2,\text{out}}-U_{3,\text{out}}&=-\frac{1}{2M_2 R_0 }\frac{\eta_\text{in} \mathfrak{p}_\text{in}}{V_{2,\text{in}}'}.\label{eq:firewall_transformation_4}
\end{align}
These four equations constitute the ADM analog of 't Hooft's firewall transformation. Note that $U_{i,\text{out}}$, $V_{i,\text{in}}$, $\mathfrak{p}_\text{out}/U_{i,\text{out}}'$, and $\mathfrak{p}_\text{in}/V_{i,\text{in}}'$  are all constants of the motion. However, with a view toward quantizatrion, we would like to be able to interpret~\eqref{eq:firewall_transformation_1} through~\eqref{eq:firewall_transformation_4} as relations between canonical quantities, not just constants of the motion.  
One possibility would be to insist that $U_{i,\text{out}}=U_{i,\text{out}}(t,\mathfrak{r}_\text{out})$ and $V_{i,\text{in}}=V_{i,\text{in}}(t,\mathfrak{r}_\text{in})$ apply at all times $t$, even outside their normal domain of region $i$. Equations~\eqref{eq:firewall_transformation_1} through~\eqref{eq:firewall_transformation_4} then define the coordinates $(U_i,V_i)$ across the shells to all regions of the spacetime. More in line with the procedure of 't Hooft would be to consider some ``global'' Kruskal coordinates $U$ and $V$, which would experience a shift at the collision point given by~\eqref{eq:firewall_transformation_1} through~\eqref{eq:firewall_transformation_4}. Regardless of interpretation, the expressions are exact and independent of foliation. Thus, the canonical pairs $(\mathfrak{r}_\text{in/out},\mathfrak{p}_\text{in/out})$ for the shells may represent any spacetime coordinate for the shells and its canonical momentum. Different choices of $t$ and $r$ will result in different definitions of $\mathfrak{p}_\text{in/out}$ and different expressions for $U'_{i,\text{out}}$ and $V'_{i,\text{in}}$.

\section{Discussion}
\label{sec:discussion}

The primary aim of this paper has been to begin bridging the gap between the literature for semiclassical and quantum models for spherical shells in a black hole background and the firewall transformation as proposed by 't Hooft. We began by generalizing the toy model used in~\citep{louko_1998} to include two spherical shells, one ingoing and one outgoing. The results unique to the two-shell case include~\eqref{eq:dd_1} through~\eqref{eq:dd_6} for the spacetime canonical variables, and~\eqref{eq:dpin_dot} and~\eqref{eq:dpout_dot} for the shells' canonical variables, which together show how all canonical variables change across both shells at the point of collision. Perhaps of more physical importance is the derivation of the Dray-'t Hooft-Redmount formula,~\eqref{eq:mass_cond} in its various forms. While the formula itself was previously derived in~\citep{dray_1985,Redmount_1985} and discussed in~\citep{Jezierski_2003,Kouletsis_2001_null_shells_3}, its various interpretations as conservation of energy equations or equations relating the momenta of the spherical shells as written here are important to the construction of the classical firewall transformation.

The final result of our paper is given in~\eqref{eq:firewall_transformation_1} through~\eqref{eq:firewall_transformation_4}, which are general classical expressions for the shift in Kruskal coordinates due to shells passing by each other.  We understand that 't Hooft wished to quantize results like these in order to form his 
firewall transformation. Our toy model of spherical shells is simplified from the case of particles, but the removal of the angular dependence does not significantly affect the important qualitative aspects of the model, including energy transfer and the Shapiro time delay. The expressions~\eqref{eq:firewall_transformation_1} through~\eqref{eq:firewall_transformation_4} are highly desirable for future quantization because they (1) include no approximation and (2) are completely coordinate independent. Previous characterizations of the shift in Kruskal coordinates have been derived in the limit that the collision occurs at the event horizon $R_0\rightarrow 2M$ as well as the limit of small shells (or particles) $E_\text{in},E_\text{out}\rightarrow 0$, whereas no such limits are required to derive~\eqref{eq:firewall_transformation_1} through~\eqref{eq:firewall_transformation_4}. Furthermore, the foliation radial coordinate $r$ has remained completely arbitrary. Thus, the canonical pair $(\mathfrak{r},\mathfrak{p})$ for the shells may take on any spacetime coordinate and its canonical momentum, which will result in different quantum theories after promoting $\mathfrak{r}_\text{in/out}$ and $\mathfrak{p}_\text{in/out}$ to quantum operators.

In his recent papers, 't Hooft has refered to the shift in Kruskal coordinates as a Shapiro time delay effect.~\citep{'thooft_2021,'thooft_2022} This terminology can be understood in the context of~\eqref{eq:T1_T4} through~\eqref{eq:T3_T4}, which show how the Schwarzschild time of the collision changes across the shells. This is a non-standard interpretation of the Shapiro effect, which normally describes the time delay in a signal as it passes by a massive body from a nearly flat region of space to another nearly flat region of space.~\citep{shapiro_1964}. In this context, the signal is one of the shells, which travels between a region of almost flat space and the event horizon of a black hole. Whether or not one can call the shift in the modified Kruskal coordinates a ``Shapiro time delay'', the shift does carry physical information about the shell moving in the opposite direction in the form of its canonical momentum.

The paper aims to make more accessible the underlying assumptions of the firewall transformation and to spark further discussion into the physical meaning of the theory as proposed by 't Hooft. We supply a generalized framework for canonical quantization, given that our classical analogs of~\eqref{eq:firewall_thooft_1} and~\eqref{eq:firewall_thooft_2}, namely~\eqref{eq:firewall_transformation_1} through~\eqref{eq:firewall_transformation_4}, require no approximation and are coordinate independent. In the context of~\citep{Vaz_2022,Vaz_2022_2}, it is clear that the final quantum theory depends on the choice of the foliation coordinates $(t,r)$, particularly the time coordinate. We leave to future work quantizations of the Hamiltonian theory proposed here, but the versions of~\eqref{eq:firewall_transformation_1} through~\eqref{eq:firewall_transformation_4} that remain after quantization will either be compatible with 't Hooft's firewall transformation, or will form a meaningful alternative. Finally, a coherence analysis similar to that in~\citep{Eyheralde_2017,Eyheralde_2019} would help determine whether or not the quantum firewall transformation actually helps to resolve the black hole information paradox.

\section*{Acknowledgements}  
NAS acknowledges support from the Graduate Student Fellowship, the CLAS Dissertation Fellowship, and a teaching assistantship from the University of Florida.  BFW acknowledges support from NSF through grant PHY-1607323, and from the University of Florida.  Hospitality at the Observatoire de Meudon, the Institut d'Astrophysique de Paris, and at the Albert Einstein Institute in Potsdam throughout the course of this work is also gratefully acknowledged.
\vfill\eject

\bibliographystyle{iopart-num}
\bibliography{references}

\providecommand{\newblock}{}
\begin{thebibliography}{10}
\expandafter\ifx\csname url\endcsname\relax
  \def\url#1{{\tt #1}}\fi
\expandafter\ifx\csname urlprefix\endcsname\relax\def\urlprefix{URL }\fi
\providecommand{\eprint}[2][]{\url{#2}}

\bibitem{Hajicek_1995}
Hajicek P 1995 {\em J. Math. Phys.\/} {\bf 36} 4612--4638
  \urlprefix\url{https://doi.org/10.1063/1.530911}

\bibitem{Hajicek_1995_2}
Hajicek P, Higuchi A and Tolar J 1995 {\em J. Math. Phys.\/} {\bf 36}
  4639--4666 \urlprefix\url{https://doi.org/10.1063/1.530912}

\bibitem{stephens_1994}
Stephens C~R, 't~Hooft G and Whiting B~F 1994 {\em Classical and Quantum
  Gravity\/} {\bf 11} 621--647
  \urlprefix\url{https://doi.org/10.1088/0264-9381/11/3/014}

\bibitem{Parentani:1994ij}
Parentani R and Piran T 1994 {\em Phys. Rev. Lett.\/} {\bf 73} 2805--2808
  (\textit{Preprint} \eprint{hep-th/9405007})

\bibitem{Ambrus_2005}
Ambrus M and Hajicek P 2005 {\em Phys. Rev. D\/} {\bf 72} 064025
  \urlprefix\url{https://doi.org/10.1103/PhysRevD.72.064025}

\bibitem{Vaz_2001}
Vaz C, Witten L and Singh T~P 2002 {\em Phys. Rev. D\/} {\bf 65} 104016
  \urlprefix\url{https://doi.org/10.1103/PhysRevD.65.104016}

\bibitem{Leal_2017}
Leal P, Bernardini A~E and Bertolami O 2018 {\em Class. Quant. Grav.\/} {\bf
  35} 115013 \urlprefix\url{https://doi.org/10.1088/1361-6382/aac083}

\bibitem{Vaz_2022_2}
Vaz C 2022 {\em Int. J. Mod. Phys. D\/} {\bf 31} 2241001
  \urlprefix\url{https://doi.org/10.1142/S0218271822410012}

\bibitem{Vaz_2022}
Vaz C 2022 {\em Phys. Rev. D\/} {\bf 105} 086020
  \urlprefix\url{https://doi.org/10.1103/PhysRevD.105.086020}

\bibitem{Hajicek_1999}
Hajicek P 2000 {\em Nucl. Phys. B Proc. Suppl.\/} {\bf 88} 114--123
  \urlprefix\url{https://doi.org/10.1016/S0920-5632(00)00759-3}

\bibitem{Hajicek_2000}
Hajicek P 2001 {\em Nucl. Phys. B\/} {\bf 603} 555--577
  \urlprefix\url{https://doi.org/10.1016/S0550-3213(01)00140-7}

\bibitem{Bicak_2003}
Bicak J and Hajicek P 2003 {\em Phys. Rev. D\/} {\bf 68} 104016
  \urlprefix\url{https://doi.org/10.1103/PhysRevD.68.104016}

\bibitem{Fiamberti_2007}
Fiamberti F and Menotti P 2008 {\em Nucl. Phys. B\/} {\bf 794} 512--537
  \urlprefix\url{https://doi.org/10.1016/j.nuclphysb.2007.11.003}

\bibitem{Corichi_2001}
Corichi A, Cruz G, Minzoni A, Padilla P, Rosenbaum M, Ryan Jr M~P, Smyth N~F
  and Vukasinac T 2002 {\em Phys. Rev. D\/} {\bf 65} 064006
  \urlprefix\url{https://doi.org/10.1103/PhysRevD.65.064006}

\bibitem{Campiglia_2016}
Campiglia M, Gambini R, Olmedo J and Pullin J 2016 {\em Class. Quant. Grav.\/}
  {\bf 33} 18LT01
  \urlprefix\url{https://doi.org/10.1088/0264-9381/33/18/18LT01}

\bibitem{Menotti_2009}
Menotti P 2010 {\em Class. Quant. Grav.\/} {\bf 27} 135008
  \urlprefix\url{https://doi.org/10.1088/0264-9381/27/13/135008}

\bibitem{Menotti_2010}
Menotti P 2010 {\em J. Phys. Conf. Ser.\/} {\bf 222} 012051
  \urlprefix\url{https://doi.org/10.1088/1742-6596/222/1/012051}

\bibitem{Eyheralde_2017}
Eyheralde R, Campiglia M, Gambini R and Pullin J 2017 {\em Class. Quant.
  Grav.\/} {\bf 34} 235015
  \urlprefix\url{https://doi.org/10.1088/1361-6382/aa8e30}

\bibitem{Eyheralde_2019}
Eyheralde R, Gambini R and Pullin J 2020 {\em Class. Quant. Grav.\/} {\bf 37}
  065001 \urlprefix\url{https://doi.org/10.1088/1361-6382/ab6e89}

\bibitem{mathur_2009}
Mathur S~D 2009 {\em Classical and Quantum Gravity\/} {\bf 26} 224001
  \urlprefix\url{https://doi.org/10.1088/0264-9381/26/22/224001}

\bibitem{'thooft_2016}
't~Hooft G 2016 {\em Found. Phys.\/} {\bf 46} 1185--1198
  \urlprefix\url{https://doi.org/10.1007/s10701-016-0014-y}

\bibitem{'thooft_2016_2}
't~Hooft G 2016 {\em Found. Phys.\/} {\bf 47} 1503--1542
  \urlprefix\url{https://doi.org/10.1007/s10701-017-0122-3}

\bibitem{'thooft_2016_3}
't~Hooft G 2016 The quantum black hole as a hydrogen atom: Microstates without
  strings attached \urlprefix\url{https://doi.org/10.48550/arxiv.1605.05119}

\bibitem{'thooft_2018}
't~Hooft G 2018 {\em Found. Phys.\/} {\bf 48} 1134 -- 1149
  \urlprefix\url{https://doi.org/10.1007/s10701-017-0133-0}

\bibitem{'thooft_2021}
’t Hooft G 2021 {\em Universe\/} {\bf 7} 298
  \urlprefix\url{https://doi.org/10.3390/universe7080298}

\bibitem{'thooft_2022}
't~Hooft G 2022 Quantum clones inside black holes
  \urlprefix\url{https://doi.org/10.48550/ARXIV.2206.04608}

\bibitem{'tHooft_2023}
t'~Hooft G 2023 {How an exact discrete symmetry can preserve black hole
  information or Turning a black hole inside out}
  \urlprefix\url{https://doi.org/10.48550/arXiv.2301.08708}

\bibitem{hogan_2020}
Hogan C 2020 {\em Class. Quant. Grav.\/} {\bf 37} 095005
  \urlprefix\url{https://doi.org/10.1088/1361-6382/ab7964}

\bibitem{zeng_2021}
Zeng D~f 2022 {\em Nucl. Phys. B\/} {\bf 977} 115722
  \urlprefix\url{https://doi.org/10.1016/j.nuclphysb.2022.115722}

\bibitem{kwon_2022}
Kwon O 2022 {Phenomenology of Holography via Quantum Coherence on Causal
  Horizons} \urlprefix\url{https://doi.org/10.48550/arXiv.2204.12080}

\bibitem{slagter_2022}
Slagter R~J 2022 {\em Int. J. Mod. Phys. A\/} {\bf 37} 2250176
  \urlprefix\url{https://doi.org/10.1142/S0217751X22501767}

\bibitem{egorov_2022}
Egorov V, Smolyakov M and Volobuev I 2023 {\em Phys. Rev. D\/} {\bf 107}(2)
  025001 \urlprefix\url{https://doi.org/10.1103/PhysRevD.107.025001}

\bibitem{arnowitt_2008}
Arnowitt R, Deser S and Misner C~W 2008 {\em General Relativity and
  Gravitation\/} {\bf 40} 1997--2027
  \urlprefix\url{https://doi.org/10.1007/s10714-008-0661-1}

\bibitem{Kuchar_1994}
Kuchar K~V 1994 {\em Phys. Rev. D\/} {\bf 50} 3961--3981
  \urlprefix\url{https://doi.org/10.1103/PhysRevD.50.3961}

\bibitem{dray_1985_2}
Dray T and 't~Hooft G 1985 {\em Commun. Math. Phys.\/} {\bf 99} 613--625
  \urlprefix\url{https://doi.org/10.1007/BF01215912}

\bibitem{Hajicek_2001_null_shells_1}
Hajicek P and Kouletsis I 2002 {\em Class. Quant. Grav.\/} {\bf 19} 2529--2550
  \urlprefix\url{https://doi.org/10.1088/0264-9381/19/10/302}

\bibitem{Hajicek_2001_null_shells_2}
Hajicek P and Kouletsis I 2002 {\em Class. Quant. Grav.\/} {\bf 19} 2551--2566
  \urlprefix\url{https://doi.org/10.1088/0264-9381/19/10/303}

\bibitem{Kouletsis_2001_null_shells_3}
Kouletsis I and Hajicek P 2002 {\em Class. Quant. Grav.\/} {\bf 19} 2567--2586
  \urlprefix\url{https://doi.org/10.1088/0264-9381/19/10/304}

\bibitem{Jezierski_2003}
Jezierski J 2003 {\em J. Math. Phys.\/} {\bf 44} 641--661
  \urlprefix\url{https://doi.org/10.1063/1.1512973}

\bibitem{louko_1998}
Louko J, Whiting B~F and Friedman J~L 1998 {\em Phys. Rev. D\/} {\bf 57}(4)
  2279--2298 \urlprefix\url{https://doi.org/10.1103/PhysRevD.57.2279}

\bibitem{Hajicek_2001}
Hajicek P and Kiefer C 2001 {\em Nucl. Phys. B\/} {\bf 603} 531--554
  \urlprefix\url{https://doi.org/10.1016/S0550-3213(01)00141-9}

\bibitem{dray_1985}
Dray T and 't~Hooft G 1985 {\em Nucl. Phys. B\/} {\bf 253} 173 -- 188
  \urlprefix\url{https://doi.org/10.1016/0550-3213(85)90525-5}

\bibitem{Redmount_1985}
Redmount I~H 1985 {\em Prog. Theor. Phys.\/} {\bf 73} 1401--1426
  \urlprefix\url{https://doi.org/10.1143/PTP.73.1401}

\bibitem{shapiro_1964}
Shapiro I~I 1964 {\em Phys. Rev. Lett.\/} {\bf 13}(26) 789--791
  \urlprefix\url{https://doi.org/10.1103/PhysRevLett.13.789}

\end{thebibliography}

\end{document}